# Technology, education and critical media literacy: potential, challenges, and opportunities

Mayte Santos Albardía*, Simón Peña-Fernández and Irati Agirreazkuenaga

Department of Journalism, Euskal Herriko Unibertsitatea/Universidad del País Vasco, Bilbao, Spain

*CORRESPONDENCE
Mayte Santos Albardía
mayte.albardia@ehu.eus

This study examines the impact of technology within media education, media literacy, and educommunication, and explores how these fields are perceived and understood by students and academic experts, focusing on the development of critical competencies and critical media literacy. Based on semi-structured in-depth interviews with leading experts in the field of critical media literacy, and a survey conducted with 141 university students in Communication and Education programs, this study explores how recent technological advances are linked to challenges in information consumption—such as disinformation, fake news, incidental exposure to information, and deepfakes—as well as the challenges and opportunities these issues present within educational contexts. The results reveal that, although such technologies provide opportunities to improve teaching–learning processes, their inclusion in the curriculum is limited and often superficial. In addition, we identify shortcomings in how teachers are trained to manage these tools effectively, hindering the development of critical thinking by students. The conclusions suggest a need for critical media literacy that not only educates students in the use of technologies but also allows them to question and evaluate the content they consume, all within an ethical and reflective framework that promotes participatory and critical citizenship.

## 1 Introduction

During the past decade, the integration of technology into education has been a catalyst for profound transformations in teaching and learning methods. Indeed, this transformation process accelerated with the COVID-19 pandemic, which resulted in significant changes in education owing to the massive adoption of digital technologies, imposing a technology-mediated pedagogy (Hernández Ortega and Cortés de las Heras, 2020).

In addition to its advantages such as interactivity and personalization of education, the rapid adoption of digital technologies has also led to new challenges. The lack of training of both teachers and students to adapt to these changes has become evident, highlighting the need to develop digital and informational competencies (Díaz Vera et al., 2021). These competencies are crucial not only for the effective use of technologies, but also to help students navigate this information-rich digital environment in a critical and reflective manner (Gutiérrez-Martín et al., 2022).

Therefore, technology not only has the potential to contribute to learning by expanding access to information but can also motivate the development of critical thinking (González-Mohíno et al., 2023). In this sense, global organizations such as UNESCO (2024) as well as authors such as Ferrés et al. (2018) and Pérez-Rodríguez et al. (2019) have stressed the importance of fostering the development of critical competencies in various environments when it comes to using information and digital technologies, through critical media literacy or educommunication.

At the same time, artificial intelligence (AI) has emerged during this same period as a tool with the potential to personalize education, and its use has already become widespread among the academic and student community (Cornejo-Plaza and Cippitani, 2023). However, for AI to have a positive impact, it must be used within an ethical and reflective framework to ensure proper training, regulation, and trust in its use (Bond et al., 2024), and to promote not only efficiency but also the development of critical and analytical thinking (Ortiz et al., 2024). The implementation of AI in classrooms must therefore also be aligned with a critical media literacy approach that enables students not only to use such technologies to their advantage but also to understand their implications and associated risks (Flores-Vivar and García-Peñalvo, 2023). In this regard, teachers have a key role to play through their teaching methods and the responsible use of these technologies within them (Sanabria-Navarro et al., 2023).

In light of the rapid growth and evolution of digital technologies and content, there is an urgent need to reconsider both the role and the implementation of critical media literacy in education. Yet, most existing studies tend to focus on theoretical frameworks or isolated interventions, without offering a comprehensive perspective that integrates expert insights with empirical data from students.

This article seeks to fill that gap by employing a mixed-methods approach to explore how future educators perceive the challenges and opportunities of media literacy within the context of technological transformation —alongside students' perceptions. By bringing together expert discourses and student experiences, this research aims to contribute to the field of educommunication by proposing an integrated, participatory, and critical framework for media education in higher education. This framework addresses pedagogical, didactic, and curricular dimensions, while also engaging with broader social issues highlighted in prior studies (Arcila Rodríguez et al., 2022).

In this context, this article poses the following research questions:

> RQ1. What is the current role of critical thinking in media and information literacy, and what role should it play in response to ongoing technological transformations?
>
> RQ2. What are the current educational challenges in the age of technological transformation?
>
> RQ3. What should be the role of technology in education and in the classroom?
>
> RQ4. How do future educators and journalists perceive their media habits and competencies in relation to critical media literacy?

## 2 Theoretical framework

The concept of educommunication was introduced in the 1980s by the popular educator Kaplún (2001), and was further developed by other influential figures such as Freire (2015), Martín-Barbero, and Prieto Castillo (Toth et al., 2012), who contributed to the construction of critical pedagogies that promoted dialog and the interrelation between education and communication as key tools for social transformation.

While many definitions have been proposed for educommunication, a concise and widely accepted definition of media literacy emerged over a decade ago as "the ability to access, analyze, evaluate and create messages in a variety of forms" (Livingstone, 2004), to which one can add its suitability for educating citizens so they can use information critically (Lopez-Gonzalez et al., 2023).

According to some experts in the field of educommunication or media literacy (Aguaded et al., 2022; Buckingham, 2020; Kellner and Share, 2019; Livingstone, 2004), the integration of technological tools, including AI, has facilitated access to information and provided other advantages (such as collaborative learning) but has also posed significant challenges, such as vulnerability to misinformation (Marta-Lazo, 2023) or fake news (Pérez Forteza and Izquierdo Cuellar, 2020), thus underlining the need for critical training for both teachers and students. In this sense, in the educational context, the ability of students to distinguish between true and false information is fundamental for their development as critical and responsible citizens (Marta-Lazo, 2023).

An additional challenge regarding the use of advanced technologies in the classroom is the management of information overload, which some authors have even described as an infodemia (Sánchez-Reina and González-Lara, 2022) or infoxication (a portmanteau of information and intoxication) (Barriga Cano, 2014; Gómez Nieto, 2016; Portugal and Aguaded, 2020). Society in general, and students in particular, are constantly bombarded by information, which can make it difficult to digest and to perceive its impact, as well as to distinguish between trustworthy and unreliable sources. Educommunication or critical media literacy becomes crucial in this context by providing students with the tools necessary to filter and evaluate information critically (González-Mohíno et al., 2023).

Some approaches to educommunication focus on digital literacy, limiting themselves to the operational use of devices without engaging with the critical analysis of content. However, beyond this, there is a need for an approach that combines media and information literacy with digital literacy, where both are included in a critical way, so that students can use technologies and media content effectively and critically during their learning processes (Mesquita-Romero et al., 2022).

Within this context of technological transformation, it is also crucial to consider how structural inequalities continue to shape students' access to media literacy. Factors such as socioeconomic status, cultural norms, and the unequal distribution of educational resources and opportunities significantly influence not only access to digital technologies but also students' ability to engage critically with media content — highlighting the persistence of the digital divide in educational contexts (Villao Salinas and Matamoros Dávalos, 2024). Research indicates that students from lower-income backgrounds and minority groups often face barriers beyond device and internet access; they also struggle to develop the media and digital literacy skills essential for participation in 21st century society (Miah, 2024). These findings emphasize the extent to which socioeconomic factors shape technological access, engagement, and educational outcomes (Joshi et al., 2024). Addressing these structural barriers is therefore key to fostering a more equitable media literacy — one that enables all students to fully and responsibly participate in digital society.

# 3 Materials and methods

This study employs a mixed-methods approach that integrates both quantitative and qualitative techniques to gain a comprehensive understanding of the topic. On one hand, semi-structured in-depth interviews were conducted with experts in media literacy and educommunication to address Research Questions 1, 2 and 3. These interviews aimed to capture insights from leading scholars regarding how critical media literacy and educommunication competencies are integrated into formal education systems, and how these experts conceptualize their relevance in addressing current educational challenges and technological transformations. On the other hand, a survey was administered to university students enrolled in education and journalism programs to collect data on their media consumption habits, perceptions, and self-assessed media literacy competencies, in response to Research Question 4. The qualitative data from expert interviews enrich the survey findings by adding interpretive depth, theoretical grounding, and pedagogical perspectives drawn from diverse cultural and institutional contexts.

The survey was conducted in person with a total of 141 participants. The questionnaires were completed by future educators and journalists—students enrolled in undergraduate programs in Journalism, Social Education, Early Childhood Education, and Primary Education at the University of the Basque Country (UPV/EHU). Data collection took place between October 2023 and March 2024 in the Faculty of Education (on the Araba, Gipuzkoa, and Bizkaia campuses) and in the Faculty of Social Sciences and Communication (Bizkaia campus). Among the respondents, 87.2% were between 18 and 21 years old. A total of 82.7% of participants were women, which can be partly attributed to the highly feminized nature of the degree programs included in the study.

On the other hand, the study also adopts a qualitative approach by using the semi-structured in-depth interview method, recognized for its suitability for research that requires detailed analysis of experiences and perceptions. In accordance with Vallés (2002) recommendations regarding qualitative interview design, the interviews were structured according to the research questions to be explored. The key issues to be addressed were determined, then participants were selected on the basis of their relevance to the study, considering the contributions they could potentially make.

The number of interviews was established by using the principle of theoretical saturation (Glaser and Strauss, 1967). This approach suggests that saturation is reached when the information collected is sufficient to meet the research objectives or when the data obtained do not provide new information that is relevant to the analysis (Guest et al., 2006). Meanwhile, the semistructured interview approach allows for flexibility to facilitate the adaptation of questions according to each interviewee's answers, thereby maximizing the depth of knowledge obtained (Kallio et al., 2016). This type of interview is especially useful when both general and specific information is sought, since it combines a basic set of questions with the possibility of exploring themes that emerge during the conversation (Galletta and Cross, 2013).

The recruitment was carried out by the researchers themselves. The interviewees were informed via e-mail about the objectives of the research and the importance of their contribution. A date and time were set for an interview lasting approximately 45 min. Prior to the interview, the confidential treatment of all the data collected during the conversation and obtained explicit consent from the interviewees to participate in the process were explained orally.

A total of seven interviews were conducted with leading professionals in the academic field who are specialists in educommunication and from different regions (Table 1). The participants were selected on the basis of both their importance in the field and the geographic diversity of their institutions. To ensure varied territorial representation, three professionals from US universities (UCLA and Temple University), one from a Latin American university (University of Lima), and three from Spanish universities (University of Huelva, University of Valladolid, and European Atlantic University) were included. The interviews were conducted online and recorded digitally. They were carried out according to the availability of the interviewees, between May 2023 and April 2024.

The interview guide was designed to elicit reflections that bridge theoretical depth with practical applications of media literacy. Experts were invited to reflect on the pedagogical implications of digital transformation, the role of educommunication and technology in the classroom, and the potential of critical media literacy to respond to contemporary educational and social challenges. The questions were open-ended and intentionally broad, allowing participants to articulate their own definitions, highlight existing gaps, and identify structural factors that influence students' engagement with media literacy (Table 2).

The guide included a core set of common questions for all participants, along with specific questions tailored to each expert's area of work (see Table 2). The questions were guided by three main objectives: (1) to explore experts' conceptualizations of educommunication and critical media literacy, (2) to identify structural and pedagogical challenges to integrating these frameworks into education; and (3) to reflect on dialogical and transformative strategies for educational praxis that could reinforce media literacy from a comparative regional perspective.

The first set of questions invited participants to define key concepts and interrogate dominant dichotomies in the field — such as technocentric versus critical paradigms. Subsequent questions delved into the political, pedagogical, and institutional dimensions of educommunication, particularly its capacity to foster critical thinking and dialogic practices within educational systems. Further questions addressed the current state of teacher training, the tension between fascination with technology and critical engagement, and the competencies required for educators to become educommunicators in an era of technological transformation. Finally, targeted questions encouraged deeper engagement with each expert's specific contributions—for example, the use of dialogical-critical method or the *Critical Media Literacy Framework*— drawing connections between theory and classroom practice.

This progression ensured that the interview captured both theoretical perspectives and concrete educational experiences, offering nuanced insights into the transformative potential of educommunication in educational and social contexts. Table 2 below outlines the core questions posed to all participants, as well as the specific questions adapted to the flow and focus of each individual conversation.

The results were analyzed using the NVIVO software, through which the data were coded for qualitative analysis with the aim of achieving an accurate and transparent depiction of the data (Welsh,

TABLE 1 Interviewees.

| Interviewee | Name | Country | Institution |
|---|---|---|---|
| I1 | Dr. Ignacio Aguaded | Spain | Professor of Education and Communication at the University of Huelva and president of *Grupo Comunicar*, a long-time supporter of media literacy in Spain |
| I2 | Dr. Jesús Bermejo Berros | Spain | Professor of Audiovisual Communication and Advertising at the University of Valladolid |
| I3 | Dr. Mónica Bonilla del Río | Spain | Professor and Researcher at the European Atlantic University, whose doctoral thesis covered media literacy and educommunication |
| I4 | Dr. Sherri Hope | United States | Professor and Researcher, and Director the Center for Media and Information Literacy (CMIL), at Temple University and Vice President of the Global Media and Information Literacy Alliance |
| I5 | Dr. Douglas Kellner | United States | Professor and Distinguished Researcher in Education at the University of California Los Angeles (UCLA) |
| I6 | Dr. Jeff Share | United States | Professor and Researcher at the University of California Los Angeles (UCLA) and faculty on the Teacher Training Program and Degrees in Education and Information Studies at UCLA |
| I7 | Dr. Julio César Mateus | Peru | Associate Professor and Researcher at the Faculty of Communication at the University of Lima. Coordinator of the Communication, Education, and Culture (CEC-IDIC) research group, and director of the *Contratexto* academic journal |

*Source: authors' own elaboration.

TABLE 2 Interview guide.

| Interviewee | Name | Common Questions | Specific Questions |
|---|---|---|---|
| I1 | Dr. Ignacio Aguaded | 1. Based on your academic and professional experience, how would you define educommunication?<br>2. How would you describe the relationship—and key differences—between educommunication, media literacy, and media education?<br>3. Many studies highlight a divide between a technocentric approach and one centered on critical media literacy. Do you find this distinction accurate? Would you suggest alternative ways to frame these perspectives?<br>4. Some authors emphasize the importance of reinforcing the dialogic dimension of educommunication to foster critical thinking. In your view, where should this shift begin—at the political level, the individual level, or within educator training?<br>5. What conditions are necessary to cultivate critical thinking in education? Do you believe educators—and their students—currently demonstrate this capacity?<br>6. Have you observed meaningful changes in teacher training over recent years, particularly in moving beyond a purely technological focus in media education?<br>7. Would you say there is a fascination with technology in education and society? Do you think this fascination influences how educommunication is practiced in certain regions?<br>8. In your opinion, what key skills distinguish an educator as an educommunicator?<br>10. Is there anything else you would like to add to our conversation? | |
| I2 | Dr. Jesús Bermejo Berros | | 11. In your work on the dialogic-critical method for fostering narrative thinking, you distinguish between conversational and critical dialog across different groups. Have you ever combined these approaches to observe before-and-after effects? How do you think this would work in practice?<br>12. Do you believe applying the dialogic-critical method in the training of future educators could help them integrate it effectively into their own teaching? |
| I3 | Dr. Mónica Bonilla del Río | | |
| I4 | Dr. Sherri Hope | | |
| I5 | Dr. Douglas Kellner | | 11. In your work *"Critical media literacy, democracy, and the reconstruction of education"* you argue that educommunication can help transform education and raise awareness of structural inequalities related to gender, race, and class. How do you see this process unfolding in practice?<br>12. Do you apply your *Critical Media Literacy Framework* in the training of future educators? If so, how? |
| I6 | Dr. Jeff Share | | 11. How would you describe the current state of educommunication in the United States? What key differences stand out when compared to other countries? Is there anything that makes the North American approach particularly distinct?<br>12. You've mentioned working with educators in training—how do you approach this work in practice?<br>13. In "*Critical media literacy, democracy, and the reconstruction of education*" you argue that educommunication can drive social transformation by raising awareness of inequalities related to gender, race, and class. How does this take shape in educational contexts?<br>14. Given the constant flow of media content, do you think it's feasible to implement dialogic, educommunication-based methodologies in teacher education? In your experience, can such approaches help future educators develop critical media literacy skills?<br>15. Do you believe educommunication should be introduced as a formal subject in undergraduate programs? |
| I7 | Dr. Julio César Mateus | | |

*Source: authors' own elaboration.

2002). Through the coding of the seven interviews that were conducted, three general ideas were identified. Together, these reveal the points of consensus and the issues where disagreement was found among the interviewees. These ideas and the subthemes within them were classified and are presented in Table 3 for schematic purposes and to facilitate their use in the "Results" section.

# 4 Results

## 4.1 Critical thinking and media and information literacy

The first general idea (GI 1), Critical thinking and media and information literacy (Table 3), highlights the relevance of media and information literacy in education, emphasizing that it requires not only technical skills but also the ability to critically analyze the media from an ethical perspective.

The results show broad consensus among the interviewed experts on the need to treat critical thinking and media and information literacy as essential competencies for navigating an environment permeated by technology and information. These findings highlight the importance of implementing these disciplines in educational settings from a critical perspective (SI 1.1), as media literacy alone may be insufficient if not supported by a solid foundation of critical thinking. For instance, in response to interview question 1— focused on the conceptualization of educommunication, Sherri Hope states:

> *"What we are all talking about is an ability to bring a critical lens to the media that we consume in all of its forms, not just digital, including books and music and everything. And that that's the intention is for us to encourage a deep reflection and an analysis of media and its influence" (personal communication, 2024).*

However, there are nuances in how the respondents defined and explained the importance of these competencies. Some experts described critical educommunication as a politically grounded pedagogical tool (SI 1.2), emphasizing its potential to strengthen democracy by addressing issues such as gender, race, class, and power. They emphasize its role in promoting participatory citizenship and social transformation, reinforcing the idea that education should not only inform, but also empower individuals to question and reshape their social realities.

Three of the seven experts interviewed highlight critical dialog as a key educommunicative tool, emphasizing the need for practical strategies rooted in dialog and critical pedagogy as core classroom practices (SI 1.3). They emphasize its value in fostering participatory learning environments where students and educators co-construct knowledge. Through dialog, learners strengthen critical thinking skills, challenge dominant narratives, and engage with diverse perspectives in a democratic educational context. As Kellner states:

> *"We give them a set of skills to critically read the media, to analyze any racism, sexism—whatever the dominant ideologies, images, or messages may be" (personal communication, 2024).*

Finally, some experts emphasize that citizens should not only consume content critically but also engage in its production ethically and consciously, underscoring the importance of fostering critical thinking throughout both processes (SI 1.4). They argue that students must not only learn to analyze media messages but also to create content consciously, understanding its potential impact and ethical implications within digital and participatory communication environments. As Bonilla (personal communication, 2024) states, there is a need of "critical consumers of content, but also producers, senders, and receivers of that information and those kind of messages".

In contrast to the emphasis placed on critical media education in academic discourse, the survey results reveal that these concepts remain relatively unknown among students enrolled in Communication and Education degree programs (Table 4). While 47.1% of respondents report having heard of media education, only 36.2% are familiar with the term media literacy. Awareness of educommunication is even lower, with just 10.9% indicating any prior knowledge of the concept. Moreover, 25.5% of participants state that they are unfamiliar with all three terms, suggesting a significant lack of exposure to these frameworks. Just 2.13% report being familiar with all three.

Although most young respondents (97.8%) recognize the importance of acquiring tools for critical media analysis, only 13.8% say these skills were addressed in their coursework. This contrast reveals a significant gap between the perceived importance of these skills and both the actual knowledge students have and the extent to which these topics are taught in their academic programs, suggesting that educational systems may not be adequately addressing media literacy in a way that reaches or engages students effectively.

## 4.2 The impact of technological transformations and current educational challenges

The second general idea (GI 2) provides evidence on the changes that technology has driven in educational environments, highlighting both the opportunities and the challenges that emerge during this process.

This highlights the omnipresence of technology and communication in daily life, which have grown substantially to the point of "getting into our bedroom, into our intimate life" (Aguaded, personal communication, 2024) and the associated need to develop critical media skills at all educational levels, including teacher training (SI 2.1). Despite the enormous educational potential of technology, the absence of a strong scientific and technological culture often leads to superficial integration—limited to the use of devices and software without a reflective and critical approach. In this regard, teachers should be prepared to guide the student body in the responsible use of such tools.

Most of the experts interviewed point out that critical media literacy remains insufficiently integrated into educational curricula (SI 2.2). They argue that, despite the growing influence of media and technology in students' lives, educational systems have yet to provide the necessary tools to navigate these environments critically. As a result, critical media literacy is often overlooked or treated superficially, leaving students exposed to the risks of media overexposure, misinformation, and uncritical content consumption. The experts emphasize the urgent need to embed critical media

TABLE 3 Insights obtained from the results.

| General idea | Secondary idea/Subtopics | Interviewee(s) |
|---|---|---|
| GI 1. Critical thinking and media and information literacy | SI 1.1 Relevance of media literacy in education from a critical perspective | I1, I2, I3, I4, I5, I6 & I7 |
| | SI 1.2 The political–pedagogical dimension of educommunication and its role in strengthening democracy | I5, I6 & I7 |
| | SI 1.3 Critical dialog as an educommunicative tool | I2, I5 & I6 |
| | SI 1.4 Content consumption and production | I1 & I3 |
| GI 2. The impact of technological transformations and current educational challenges | SI 2.1 Lack of teacher training | I1, I4, I5 & I7 |
| | SI 2.2 Educommunication in curricula | I1, I4, I5 & I7 |
| | SI 2.3 Social factors | I4 |
| GI 3. Use and management of the internet in the classroom | SI 3.1 Fake news, disinformation, and incidentally received information | I1, I2, I3, I4, I5, I6, & I7 |
| | SI 3.2 Tools for managing information-related challenges | I2 & I6 |
| | SI 3.3 Educommunication, ideology, and power | I4 & I6 |
| | SI 3.4. Deep fakes, a new level of complexity | I3 & I5 |
| | SI 3.5. The AI educational divide | I4 & I7 |

*Source: authors' own elaboration.

TABLE 4 Student perceptions of media education in communication and education degree programs.

| | Women | | Men | | Total | |
|---|---|---|---|---|---|---|
| | N. | % | N. | % | N. | % |
| Question: About your relationship with educommunication/media literacy/media education during your degree: | | | | | | |
| I'm familiar with media literacy | 43 | 36.8 | 7 | 33.3 | 50 | 36.2 |
| I'm familiar with educommunication | 11 | 9.4 | 4 | 19.0 | 15 | 10.9 |
| I'm familiar with media education | 54 | 46.2 | 11 | 52.4 | 65 | 47.1 |
| These skills were addressed in my coursework | 17 | 14.5 | 2 | 9.5 | 19 | 13.8 |
| These skills should be part of my education | 115 | 98.3 | 20 | 95.2 | 135 | 97.8 |

*Source: authors' own elaboration.

literacy as a core component of the curriculum to foster informed and reflective digital citizens.

Additionally, one of the experts offers a complementary perspective, arguing that the potential impacts of technology are not determined solely by training, but also vary depending on broader social factors (SI 2.3). In contexts where individuals lack strong support networks—such as friends or family—or face limited financial resources, media and technologies may end up filling emotional or social voids. This dynamic increases the risk of excessive exposure or reliance, potentially leading to an overuse that surpasses what is pedagogically or socially appropriate:

> *"Media is a huge influence. But it's not the only influence. It just has the imbalance of it is that to the extent that you are missing any of those other things, so maybe you do not have a lot of friends or your parents are divorced or whatever it might be, you are low income, media will fill it in" (Sherri Hope, personal communication, 2024).*

Moreover, the expert highlights the need to consider these disparities when addressing technology integration in education.

The analysis of media consumption among university students shows that social media (77.5%) and television (63%) are the primary sources of information, followed by digital newspapers (47.1%) and general internet sources (42.8%) (Table 5). At the other end, only 6.5% listen to the radio and 8.7% read print newspapers, indicating a clear preference among young people for digital environments over traditional media. The data also reflect notable differences in the diversity of media sources used by respondents. Notably, 48.94% indicate that they use three or more sources to stay informed, compared to 17.73% who report relying on only one source of information (Table 6).

As for the frequency of news consumption, a total of 72.5% of respondents report accessing news daily or almost daily, compared to 24.6% who do so occasionally and 2.9% who never or almost never engage with news. These findings show that news consumption is a frequent habit for almost three out of four surveyed students, likely influenced by technological accessibility and the integration of media into everyday life.

When asked whether they believe they have the necessary tools to critically manage media information, slightly less than half of the respondents (45.4%) answered affirmatively (Table 5). Similarly, regarding information overload, 44.7% of students report feeling that they have the necessary tools to manage the amount of information they receive daily through the media, and only 45.7% stated that they feel ready to work with media content in their future professional roles.

This gap between the high perceived importance of acquiring critical media skills and the relatively low self-assessed competence

TABLE 5 Students' media use and frequency.

| | Women | | Men | | Total | |
|---|---|---|---|---|---|---|
| | N. | % | N. | % | N. | % |
| **Type of media consumed** | | | | | | |
| I get my news from TV | 75 | 64.1 | 12 | 57.1 | 87 | 63.0 |
| I get my news from radio | 5 | 4.3 | 4 | 19.0 | 9 | 6.5 |
| I get my news from newspapers | 9 | 7.7 | 3 | 14.3 | 12 | 8.7 |
| I get my news from online newspapers | 53 | 45.3 | 12 | 57.1 | 65 | 47.1 |
| I get my news from social media | 91 | 77.8 | 16 | 76.2 | 107 | 77.5 |
| I get my news from podcasts | 21 | 17.9 | 4 | 19.0 | 25 | 18.1 |
| I get my news from the internet (forums, blogs.) | 51 | 43.6 | 8 | 38.1 | 59 | 42.8 |
| **News consumption frequency** | | | | | | |
| Daily | 52 | 44.4 | 12 | 57.1 | 64 | 46.4 |
| Almost daily (2/3 times per week) | 29 | 24.8 | 7 | 33.3 | 36 | 26.1 |
| Weekly | 17 | 14.5 | 1 | 4.8 | 18 | 13.0 |
| Occasionally (every 2 weeks) | 15 | 12.8 | 1 | 4.8 | 16 | 11.6 |
| Monthly | 1 | 0.9 | 0 | 0 | 1 | 0.7 |
| I do not consume media | 3 | 2.6 | 0 | 0 | 3 | 2.2 |
| **Perceived Media Literacy Competence** | | | | | | |
| I can manage the amount of information I get from the media | 46 | 39.3 | 9 | 42.9 | 55 | 39.9 |
| I have the skills I need to critically analyze media information | 55 | 47.0 | 7 | 33.3 | 62 | 44.9 |
| I feel ready to work with media content | 52 | 44.4 | 11 | 52.4 | 63 | 45.7 |

*Source: authors' own elaboration.

TABLE 6 Impact of the media on opinion formation.

| | Men | | Women | | Total | |
|---|---|---|---|---|---|---|
| | N. | % | N. | % | N. | % |
| **Question: The media influence the formation of your opinion in a:** | | | | | | |
| Very positive way | 6 | 5.1 | 1 | 4.8 | 7 | 5.1 |
| Somewhat positive way | 46 | 39.3 | 9 | 42.9 | 55 | 39.9 |
| Somewhat negative way | 62 | 53.0 | 10 | 47.6 | 72 | 52.2 |
| Very negative way | 3 | 2.6 | 1 | 4.8 | 4 | 2.9 |

*Source: authors' own elaboration.

highlights a pressing challenge for educational systems. This suggests that although students recognize the value of media literacy, they often lack the strategies to filter, prioritize, and critically process the overwhelming volume of information they face—highlighting the need for more effective and accessible training in critical media analysis.

Cross-referencing the survey data with the frequency of media consumption reveals several interesting patterns, shedding light on additional factors that contribute to a more comprehensive understanding of the phenomenon. A correlation can be observed between more frequent media consumption and a greater perceived ability to manage and critically analyze the information received. Those who consume information frequently (daily or almost daily) report a higher capacity to manage the amount of information they receive (48.5%) compared to those who consume information occasionally (40.6%). Similarly, frequent consumers are more likely (48.5%) to report having the ability to critically analyze media content than occasional consumers (34.4%).

The data also reveal a relationship between the number of media sources consumed and the perceived ability to critically analyze information. As the number of sources used for staying informed increases, so does the percentage of individuals who consider themselves equipped with the tools to engage critically with media content. While confidence in critical skills hovers around 40% among those who consume only one or two sources, it rises to over 50% among those who use four or more. Absolute frequency reinforces this trend: although fewer respondents fall into the higher-consumption categories, they are proportionally more likely to express confidence in their critical media competence.

These data suggest not only a perceived need to strengthen this learning, but also that media literacy and critical consumption skills may be developed through meaningful media engagement outside formal educational settings. While formal education is essential, students also develop critical skills through active engagement with media in their everyday lives. Encouraging diverse and intentional media use outside the classroom reinforces what is taught within it. The data suggest that media literacy can be approached both as a pedagogical objective and as an experiential process grounded in real-world media interaction.

## 4.3 Use and management of the internet in the classroom

The third general idea (GI 3) deals with the expansion of technology, in particular the internet and social media, and how to manage its integration into the classroom, a topic that was widely debated by the interviewees.

There is a general consensus among the interviewed experts regarding the growing spread of fake news, disinformation, and incidentally received information (SI 3.1). They stress that students are frequently exposed to unreliable content, especially through social media and algorithm-driven platforms, making it increasingly difficult to distinguish between credible and misleading sources. This reality underscores the urgent need to equip students with strong critical thinking and media literacy skills. Developing the ability to verify information, question sources, and reflect on content is considered essential for promoting informed, responsible, and active digital citizenship in today's complex media environment. In this regard, Aguaded affirms:

> *"For communication, the most important thing is the vaccine — that is, prevention. And prevention is called education: education for citizenship in all its contexts" (personal communication, 2024).*

Therefore, some of the experts interviewed point out that strategies such as the use of the Critical Media Literacy Framework (Kellner and Share, 2019), the dialogic–critical method in educommunication (Bermejo Berros, 2021), and critical pedagogies in the classroom are key approaches in this regard (SI 3.2). These tools allow students to identify, question, and evaluate the veracity of information and understand the context in which it is produced, thereby promoting deeper analysis of digital content and incorporating dialog as an educommunicative tool. When asked about teaching training tools applied with students, Bonilla said:

> *"There is a need for practices that make them reflect — practices that help them think about why this happens, who is behind this type of news, what their intentions might be, what ideology the media may be trying to convey, who writes the news, what source provides it, and how to truly compare it" (personal communication, 2024).*

These experts highlight the need for such educommunicative tools to address not only how to analyze a news item but also how and why such news reaches the student body, which is essential to understand dynamics such as clickbait and incidentally received information.

In addition, several experts argue that the absence of critical media literacy in education not only facilitates the spread of misinformation and fake news, but also contributes to the growing polarization seen in today's society (SI3.3). They emphasize that without the ability to critically engage with media content, students are more likely to adopt unexamined viewpoints and become entrenched in ideological bubbles. Sherri Hope affirms:

> *"Much of the polarization and misinformation we see around the world is a direct result of the lack of media literacy in education." (personal communication, 2024).*

Educommunication is seen as essential for uncovering and analyzing the power dynamics that shape media narratives, influence public discourse, and reinforce existing social inequalities, thus offering tools to resist manipulation and promote democratic dialog (SI 3.3).

Finally, the last two subtopics are related to the risks associated with artificial intelligence, the latest guest in the ongoing process of technological and digital transformation. Experts highlight that while AI offers potential benefits, its rapid and often uncritical integration into education raises serious ethical, pedagogical, and cognitive concerns.

Some experts claim that AI has introduced a new level of complexity into the digital environment, particularly through phenomena such as deep fakes. Such fakes make it even more difficult to distinguish between reality and misinformation, resulting in significant challenges for education (SI 3.4). In this regard, critical educommunication can help recognize the limitations and risks of technologies, including AI, and ensuring that it can be used appropriately by the student body is another teaching task:

> *"Being surrounded by technology means we necessarily have to develop media competencies to use it properly [.] The teacher can [transform it into] a tool that can be used appropriately and responsibly — rather than inappropriately" (Bonilla, personal communication, 2024).*

In contrast, criticism of the lack of systematic and structural integration of media language and emerging technologies such as AI into educational curricula is also expressed (SI 3.5). Regarding interview question 8, which focuses on the technological aspects of educommunication in education, Mateus states:

> *"Media language is not taught — let alone other issues, like the more ideological dimensions linked to representations, stereotypes, the role of advertising, and now the role of artificial intelligence" (personal communication, 2024).*

The final idea expressed concerns the dependence on imported rather than home-grown technologies, as well as the technological—and at times superficial—fascination of some teachers who, without proper training, begin using tools without fully understanding how they work. This shortcoming limits the effective integration of technologies into education and prevents both teachers and students from using AI in a critical, informed, and pedagogically meaningful way.

In this regard, the survey data indicate that more than half of the students surveyed (52.2%) believe that the media have a rather negative influence on the formation of their opinions. A smaller portion (39.9%) perceive the media's influence as rather positive, while more extreme views—very positive (5.1%) and very negative (2.9%)—are held by a minority. These results suggest a generally critical stance among young people toward media influence.

Among the respondents who get their news from radio (66.7%) and podcasts (64%), the influence of the media on opinion formation is perceived as positive (either very or somewhat), making these the formats with the most favorable perception. The more personalized and audio-based nature of these formats may be associated with higher levels of trust in the content received. Similarly, print media consumption is also linked to a predominantly positive perception: 58.4% of print newspaper readers view the media's role in shaping opinions favorably. In contrast, the perception is less favorable among digital press consumers, 48.5% of whom hold a positive view. These findings suggest that more traditional formats—often perceived as more rigorous—continue to inspire trust among a portion of the public. In contrast, television news consumers show a less favorable perception, with 44.3% viewing the media's influence positively. While digital platforms are widely used, trust in them appears to be lower compared to audio or print sources. Finally, among those who get their information from social media and general internet sources (such as forums and blogs), positive perception drops below half, with 45.8 and 38.3%, respectively. This may reflect greater critical awareness of the open, fast-paced, and sometimes unreliable nature of these environments.

# 5 Discussion

The results of this study coincide with recent research highlighting the importance of media literacy in the digital educational context or focusing on the development of critical thinking (Bulger and Davison, 2018; Escribano-Muñoz et al., 2024; Lopez-Gonzalez et al., 2023). Without such a solid critical base, students cannot adequately face the information overload, disinformation, and hyperdigitization of the digital sphere (García-Ruiz et al., 2020).

Integrating these critical media competencies into educational curricula should be a priority. However, the main challenges include a lack of institutional support and insufficient teacher training to effectively convey these skills to students (Moreno-Gil, 2024). In this regard, political will, along with the activism of engaged educators interested in the subject, civil society and academia, are key to promoting the integration of critical media literacy into formal education (Rojas-Estrada et al., 2024). Furthermore, this work also highlights the need to broaden the concept of media literacy to include a critical perspective that considers the economic, ideological, and cultural dimensions of media and technologies and their power to reinforce democracy (Share et al., 2019).

At the same time, it is essential to recognize how cultural, economic, and geopolitical contexts shape digital inequality and exclusion (Andrade-Vargas et al., 2021). Although internet use has become widespread, this does not imply that social inequalities have disappeared (Micheli, 2016) — on the contrary, many have been reproduced — or even deepened — within the digital sphere. The COVID-19 pandemic further intensified these digital divides, which reflect underlying structural inequalities that disproportionately affect low-income populations (Tripathi, 2024) and rural communities (Morales Romo, 2017). These disparities restrict access to knowledge and limit future employment opportunities (Agudelo Ramírez and Zuluaga Cruz, 2022). Moreover, the ongoing digital gap between the Global North and South risks exacerbating inequalities in educational outcomes on an international scale (Seoane, 2025).

These structural and digital disparities also pose significant barriers to learners' ability to become active and critical media users. Critical thinking skills can only be meaningfully developed when students have the opportunities to engage with media not only as consumers but also as producers. In this context, ensuring access to technological tools and resources becomes a fundamental step. Addressing issues related to connectivity, availability, and accessibility is essential for enabling knowledge appropriation (Terreni et al., 2017) and fostering critical thinking. This, in turn, underscores the importance of treating digital inequity as a central concern in any educational strategy (Meng et al., 2024). It also supports Buckingham's (2019) argument that media literacy must be approached through an inclusive and structural perspective — one that acknowledges the social, economic, and ideological factors shaping access, participation, and agency.

Given these disparities, media literacy education must move beyond a one-size-fits-all model and be grounded in the specific financial, cultural, and regional contexts of learners. When developed through inclusive, context-sensitive frameworks, media literacy can serve as a powerful tool to ensure that individuals from diverse backgrounds have equitable opportunities to critically engage with — and benefit from — media content and narratives (Bozdağ, 2022).

Likewise, recent studies have pointed out the need for educators to develop critical media competencies that go beyond the simple, instrumental use of technology (Osuna-Acedo et al., 2018; Sánchez et al., 2024). As mentioned in the "Results" section, the integration of technologies into education is often superficial owing to a lack of deep technological culture and institutional support for teachers (Cruz, 2019). This reflects a broader trend in which technology is used without sufficient pedagogical grounding, as many teachers lack both technical understanding and media literacy training — largely due to institutional priorities that favor technological and instructional skills over media education (Gutiérrez-Martín et al., 2022).

Artificial intelligence, with its capacity to process large amounts of data and adapt to individual needs, is considered a promising tool to enhance educational quality and optimize teaching–learning processes (Ríos Hernández et al., 2024). In fact, AI-based methods have proven effective in educational settings for identifying fake news, helping reduce the impact of misinformation on the public (Chiang et al., 2022). In this regard, the use of technology must be combined with transparency, critical media education, and regulation to mitigate the increase of misleading content and its effects (Gómez-De-Ágreda et al., 2021).

Given its potential for both positive and negative impact, artificial intelligence must be approached with awareness and critical reflection, including the definition of ethical guidelines and a rethinking of assessment methods (Mateus et al., 2024), to ensure that it does not become a tool for control or disinformation. The success of this will depend on how it is integrated into pedagogical processes and its combination with social and human competencies, which are key to fostering a balanced coexistence between technological advancement and critical-reflective practices in education.

Finally, this study has certain limitations that point to potential directions for future research. While the qualitative component draws on interviews with international experts, all of them were based in Spain, Latin America, and the United States. Incorporating perspectives from additional global regions would enrich the comparative understanding of how critical media literacy is conceptualized and applied across different contexts. Similarly, the survey data were collected exclusively in the Basque Country, which limits the generalizability of the findings. Future research could broaden the empirical scope by including student surveys across a wider range of national and regional settings, offering a more comprehensive view of how media literacy is perceived and practiced within diverse educational systems.

Moreover, the study's cross-sectional design provides only a snapshot of current perceptions, without capturing how these views and competencies evolve over time. Longitudinal research could offer valuable insights into how awareness and skills develop in response to ongoing technological change. Finally, future studies should prioritize the creation of robust and inclusive assessment tools that measure not only proficiency but also the ethical, critical, and reflective dimensions of media engagement—elements that are essential to a truly transformative media literacy.

# 6 Conclusion

Although university students generally acknowledge the relevance of acquiring media-related skills, these competencies are rarely addressed in depth throughout their academic training. Concepts such as media literacy, media education, or educommunication remain unfamiliar to many, and are often absent or only superficially introduced in coursework. Meanwhile, students tend to consume news frequently, primarily through television and social media, combining traditional and digital sources. Despite this regular exposure, many do not feel adequately prepared to critically analyze media content or manage the overwhelming volume of information they encounter daily. However, there appears to be a connection between more frequent media consumption and greater confidence in one's critical abilities, suggesting that active engagement with information sources may help strengthen media-related competencies. When reflecting on the influence of media on their opinions, students tend to express mixed views, with traditional outlets often inspiring more trust than digital platforms. These patterns point to a disconnection between the perceived importance of these skills and their actual development within university curricula. They also offer a clear direction for improvement: the need to revisit educational programs in Communication and Education so that students are not only exposed to media, but also empowered to understand, evaluate, and engage with it critically and responsibly.

The results of this study highlight the importance of and urgent need for critical media literacy in the current educational context, as the associated skills are essential for effectively and safely navigating a digital environment saturated with technology and information. However, media literacy must not be limited to the use of technological devices and digital tools but must always incorporate criticality as an educational practice.

On the other hand, critical media literacy should allow not only the ethical and reflective consumption of content but also its conscious production. In this regard, it is essential to structurally and transversally integrate this competency into educational curricula, since its absence can expose students to vulnerability and even risk owing to their constant overexposure to advanced technology and digital information, which may not always be accurate.

This lack of integration of a key competency not only facilitates the spread of disinformation and fake news but can also contribute to social polarization. Therefore, educommunicative tools, such as the Critical Media Literacy Framework, the dialogic–critical method in educommunication, and critical pedagogies, are essential to construct a critical population, to promote democracy, and to build an education system that is at the service of values and capable of providing students with the skills required to analyze, question, and evaluate information in increasingly complex contexts.

Although digital tools have the potential to transform and improve teaching–learning processes, the lack of a deep scientific and technological culture often leads to a superficial implementation of technology in educational environments. Sometimes, both teachers and students limit their interaction with technology to an operational use but without a reflective approach capable of fostering a critical and responsible use that genuinely benefits the educational system at a structural level. In addition, the expansion of technologies such as artificial intelligence adds a new level of complexity to their use and implementation in the educational environment, requiring specific, critical, and adequate training on their potential, limitations, benefits, and risks.

In this context, it is really important to develop critical media skills at all educational levels, including teacher training, and to develop evaluation tools that assess not only technical proficiency but also critical media competencies. Educators should implement assessments that measure ethical reasoning, analytical skills, and students' ability to engage in reflexive media practices.

Nonetheless, it is essential to acknowledge that the implementation and impact of media literacy education are profoundly shaped by the cultural, regional, and structural contexts in which they take place. In countries and regions marked by inequality, limited access to technological infrastructure and educational opportunities creates additional obstacles to developing critical media skills. These disparities are not merely technical—they are deeply social and political, calling for context-sensitive strategies that address the structural roots of exclusion. Therefore, advancing critical media literacy must go beyond pedagogical innovation: it requires structural commitments to equity in access, participation, and representation within both media and education systems.

To sum up, this article contributes to the field of media literacy and digital education by exposing a persistent gap between the importance that both students and experts attribute to media literacy and its limited integration into educational programs. The study identifies curricular gaps and underscores the need for a more critical, participatory, and inclusive approach to media literacy pedagogy. By incorporating diverse regional perspectives and emphasizing critical reflection, social justice, and technological ethics, this article offers both empirical evidence and a conceptual framework to support educators and policymakers in rethinking media education in the context of digital transformation.

# Data availability statement

The original contributions presented in the study are included in the article/supplementary material, further inquiries can be directed to the corresponding author/s.

## Ethics statement

The studies involving humans were approved by CEISH-UPV/EHU (Universidad del País Vasco) (protocol code M10_2023_147 on July 21, 2023). The studies were conducted in accordance with the local legislation and institutional requirements. The participants provided their written informed consent to participate in this study.

## Author contributions

MSA: Methodology, Conceptualization, Writing – original draft, Validation, Formal analysis, Resources, Data curation. SP-F: Writing – review & editing, Supervision, Validation, Methodology. IA: Writing – review & editing, Validation, Methodology, Supervision.

## Funding

The author(s) declare that no financial support was received for the research and/or publication of this article.

## Conflict of interest

The authors declare that the research was conducted in the absence of any commercial or financial relationships that could be construed as a potential conflict of interest.

## Generative AI statement

The authors declare that no Gen AI was used in the creation of this manuscript.

## Publisher's note

All claims expressed in this article are solely those of the authors and do not necessarily represent those of their affiliated organizations, or those of the publisher, the editors and the reviewers. Any product that may be evaluated in this article, or claim that may be made by its manufacturer, is not guaranteed or endorsed by the publisher.